\documentclass{elsart}
\usepackage{graphicx}

\def\eqref#1{eq.~(\ref{#1})}
\def\figref#1{Fig.~\ref{#1}}

\begin{document}

\begin{frontmatter}

\title{Do Pareto-Zipf and Gibrat laws hold true?\\
  An analysis with European Firms}
\author[ATR]{Yoshi Fujiwara\thanksref{contact}}
\ead{yfujiwar@atr.co.jp}
\author[Ancona]{Corrado Di Guilmi}
\author[Kyoto]{Hideaki Aoyama}
\author[Ancona]{Mauro Gallegati}
\author[ATR]{Wataru Souma}

\address[ATR]{%
  ATR Human Information Science Laboratories,
  Kyoto 619-0288, Japan}
\address[Ancona]{%
  Department of Economics,
  Universit\'a Politecnica delle Marche,
  P. Martelli 8, I-62100 Ancona, Italy}
\address[Kyoto]{%
  Graduate School of Science, Kyoto University,
  Kyoto 606-8501, Japan}
\thanks[contact]{%
  Corresponding author. FAX: +81-774-95-2647.
}

\begin{abstract}
  By employing exhaustive lists of large firms in European countries,
  we show that the upper-tail of the distribution of firm size can be
  fitted with a power-law (Pareto-Zipf law), and that in this region
  the growth rate of each firm is independent of the firm's size
  (Gibrat's law of proportionate effect). We also find that detailed
  balance holds in the large-size region for periods we investigated;
  the empirical probability for a firm to change its size from a value
  to another is statistically the same as that for its reverse
  process. We prove several relationships among Pareto-Zipf's law,
  Gibrat's law and the condition of detailed balance. As a
  consequence, we show that the distribution of growth rate possesses
  a non-trivial relation between the positive side of the distribution
  and the negative side, through the value of Pareto index, as is
  confirmed empirically.
\end{abstract}



\end{frontmatter}


\section{Introduction}

Pareto \cite{Pareto} is generally credited with the discovery, more
than a century ago, that the distribution of personal income obeys a
{\it power-law\/} in high-income range\footnote{%
  See \cite{Aoyama00} for modern and high-quality personal-income data
  in Japan.}%
. Firm size also has a skew
distribution \cite{SKEW}, and quite often obeys a power-law in the
upper tail of the distribution. In terms of cumulative distribution
$P_>(x)$ for firm size $x$, this states that
\begin{equation}
  P_>(x) \propto x^{-\mu} ,
\label{pareto}
\end{equation}
for large $x$, with $\mu$ being a parameter called Pareto index. The
special case $\mu=1$ is often referred to as Zipf's law \cite{Zipf}. In
this paper we call it Pareto-Zipf law, the fact that firm size has a
power-law distribution asymptotically for large firms.

Even if the range for which \eqref{pareto} is valid is a few percent
in the upper tail of the distribution, it is often observed that such
a small fraction of firms occupies a large amount of total sum of firm
sizes.  This means that a small idiosyncratic shock can make a
considerable macro-economic impact. It is, therefore, quite important
to ask what is the underlying dynamics that governs the growth of
those large firms.

Let a firm's
size be $x_1$ at a time and $x_2$ at a later time. Growth rate $R$ is
defined as the ratio $R=x_2/x_1$. {\it Law of proportionate effect}
\cite{Gibrat} (see also \cite{Steindl}) is a postulate that the growth
rate of a firm is independent of the firm's attained size, i.e.
\begin{equation}
  P(R|x_1) \textrm{ is independent of } x_1 ,
  \label{gibrat}
\end{equation}
where $P(R|x)$ is the probability distribution of growth rate
conditional on the initial size $x_1$. In this paper we call this
assumption as Gibrat's law\footnote{%
  Another interesting and related quantity is {\it flow},
  e.g.~profits, rather than stock. See \cite{Yoshi03a} for growth of
  individual personal-income and \cite{Aoyama03a} for firms tax-income
  growth, and validity of Gibrat's law.}%
.

These two laws have been extensively studied in industrial
organization and related stochastic models
\cite{SKEW,Steindl,Kalecki45,Champer53,Simon55,Hart56,Hymer62,Mansfield62,Samuels65,Singh75,Chesher79,Hall87,Evans87,Dunne88,Sutton95,McLoughan95,Hart96,Hart97}
(see \cite{Sutton97} for review).
Recent study in econophysics
\cite{Stanley95,Stanley96,Amaral97,Buldyrev97,Amaral98,Takayasu98,Okuyama99,Ramsden00,Axtell01,Mizuno02,Gaffeo03,Bottazzi03}
introduced some notions and concepts of statistical physics into
economics (see \cite{Stanley00}).
Present status related to firm-size growth may be summarized as
follows. Firm size distribution is approximately log-normal with
deviation from it in the upper tail of the distribution (e.g.
\cite{Hart97} for recent data). On the other hand, Gibrat's law breaks
down in the sense that the fluctuations of growth rate scale as a
power-law with firm size; smaller firms can possibly have larger
fluctuations (e.g. \cite{Stanley96,Amaral97}). However, little
attention has been paid to the regime of firm size where power-law is
dominant rather than log-normality, and to the validity of Gibrat's
law in that regime. More importantly, any {\it kinematic\/}
relationship between Pareto-Zipf and Gibrat laws has not been
understood explicitly, although there have been a lot of works on
stochastic dynamics since Gibrat. This issue is exactly what the
present paper addresses.

For our purpose it is crucial to employ exhaustive lists of large
firms. Our dataset for European countries is exhaustive in the sense
that each list includes all the active firms in each country whose
sizes exceed a certain threshold of observation. We show that both of
the Pareto-Zipf law and Gibrat's law do hold for those large firms. As
our main result, we prove that Pareto-Zipf law implies Gibrat's law
and {\it vice versa\/} under detailed balance. By showing that the
condition of detailed balance also holds in our empirical data, we can
show the equivalence of Pareto-Zipf law and Gibrat's law as a {\it
  kinematic\/} principle in firms growth, irrespective of the
underlying dynamics. Thereby, we conjecture that Gibrat's law does
hold in the regime of Pareto-Zipf for large firms, but does not for
smaller firms. Thus our result is not contradictory to the breakdown
of Gibrat's law in previous study, most notably to the recent work by
Stanley's group \cite{Stanley96,Amaral97,Buldyrev97,Amaral98}.
Furthermore, in the process of our proof, we also show that the
distribution of growth rate possesses a non-trivial relation between
the positive side ($R>1$) of the distribution and the negative side
($R<1$), through the value of Pareto index $\mu$, which is confirmed
empirically.

In section 2, we give a brief review of the study on Gibrat's law and
firm size distribution in economics. In section 3, we describe the
nature of our database of firms with large size in European countries.
In section 4, using exhaustive lists of large firms in the dataset, we
show that Gibrat's law holds in the power-law regime for which the
firm size distribution obeys Pareto-Zipf law. In addition, we uncover
that temporal change of individual firm's size in successive years
satisfies what we call time-reversal symmetry, or detailed balance.
In section 5, we prove that the two empirical laws of Gibrat and
Pareto-Zipf are equivalent under the condition of detailed balance.
We summarize our results in section 6.

\section{Gibrat and Pareto-Zipf Laws in Economics}

Industrial organization literature has long been focused on two
empirical facts
\cite{SKEW,Steindl,Kalecki45,Champer53,Simon55,Hart56,Hymer62,Mansfield62,Samuels65,Singh75,Chesher79,Hall87,Evans87,Dunne88,Sutton95,McLoughan95,Hart96,Hart97}
(see \cite{Sutton97} for review):\\
(i) skew distribution of firms size\\
(ii) validity or invalidity of Gibrat's law for firm growth

Gibrat formulated the law of proportionate effect for growth rate to
explain the empirically observed distribution of firms. The law of
proportionate effect states that the expected increment to a firm's
size in each period is proportional to the current size of the firm.
Let $x_t$ and $x_{t-\Delta t}$ be, respectively, the size of a firm at
time $t$ and $t-\Delta t$, and $\epsilon_t$ denote the proportionate
rate of growth. The the postulate is expressed as
\[
  x_t-x_{t-\Delta t}=\epsilon_t x_{t-\Delta t} .
\]
Gibrat assumed (a) that $\epsilon_t$ is independent of $x_t$ (Gibrat's
law), (b) that $\epsilon_t$ has no temporal correlation, and (c) that
there is no interaction between firms. Then, after a sufficiently long
time $t\gg\Delta t$, since
\[ x_t=x_0(1+\epsilon_1)(1+\epsilon_2)\cdots(1+\epsilon_t) , \]
$\log x_t$ follows a random walk. Assuming that $\epsilon_t$ is small,
one has
\[ \log x_t=\log x_0+\epsilon_1+\epsilon_2+\cdots+\epsilon_t . \]
Gibrat's model has two consequences concerning the above points (i)
and (ii). Since the growth rate defined by $R_t\equiv x_t/x_0$ has its
logarithm as the sum of independent variables $\epsilon_t$, the growth
rate is log-normally distributed. In addition, assuming that all the
firms have approximately the same starting time and size, the
distribution of firms size is also also log-normal with mean and
variance given by $mt$ and $\sigma^2 t$, respectively, where $m$ is
the mean of $\epsilon_t$ and $\sigma^2$  is the variance of
$\epsilon_t$.

The assumptions (a)--(c) in Gibrat's model are in disagreement with
empirical evidence. Among others, the Gibrat's law (a) is incompatible
with the fact that the fluctuations of growth rate measured by
standard deviation decreases as firm size increases
\cite{Hart56,Hymer62,Singh75,Hall87,Evans87}.
Especially, the recent work \cite{Stanley96,Amaral97} by Stanley's
group showed that the distribution of the logarithm of growth rates,
for each class of firms with approximately the same size, displays an
exponential form (Laplace distribution) rather than log-normal. They
also show that the fluctuations in the growth rates characterized by
the standard deviation $\sigma(x)$ of the distribution decreases for
larger size of firms as a power-law, $\sigma(x)\sim x^{-\beta}$, with
the exponent $\beta$ is less than a half. The latter point suggests a
new viewpoint about the interplay of different parts of a firm, an
industrial sector, or an organization \cite{Buldyrev97,Amaral98}.

In contrast to the standard deviation, the measure by mean growth
rate has been disputed.  There were studies which showed that smaller
firms grow faster \cite{Evans87} or slower \cite{Samuels65,Dunne88}
than bigger ones.  However, it is generally thought that the
proportional rate of growth of a firm (conditional on survival) is
decreasing in size, as far as small and medium firms are concerned,
which share a large fraction of industrial sectors in number.
However, the remaining larger firms constitute a small fraction in
number, but occupy a large fraction of total sum of firms size. This
is due to the effect of heavy tail, much heavier than expected from
log-normal regime. See recent works
\cite{Stanley95,Hart97}\footnote{%
  \cite{Stanley95} observed that log-normal distribution {\it
    overestimates} the upper-tail of size distribution based on
  Computat in U.S. As noted in the paper, the dataset is consisting of
  only publicly-traded firms. This can be a possible cause of their
  observation. \cite{Hart97} used much larger dataset in U.K. Though
  their plot showed a power-law regime over several orders of
  magnitude, they rejected the hypothesis of power-law due to the
  presence of super-giant firms. We consider that both of the these points
  deserve further investigation.}%
. This is the Pareto-Zipf regime which we focus on in this paper.

On the other hand, the assumption (b) about temporal correlation
between successive growth rates are not investigated with definite
conclusions. \cite{Chesher79}, for example, showed that the
distribution of growth rates shows a first-order positive
autocorrelation: the growth process will result faster for firms which
recorded a sharp growth in previous years. (\cite{Chesher79} also
furnished a test test for the validity of Gibrat's Law that takes into
account the ``historical memory" of the growth process.)

Gibrat's work also opened up a stream of theoretical models and
ideas. Kalecki noted that Gibrat's model leads to ``unrealistic''
feature, that is, the variance of the size distribution would increase
indefinitely with time. He considered several models, one of which
assumed that the expected rate of growth increased less than
proportionately, leading to a log-normal distribution with constant
variance.

Herbert Simon considered it more important that the firm size
distribution has heavy tail in upper-region of size, which was better
fitted by Yule distribution or asymptotically a Pareto-Zipf law. In
order to explain such a distribution, based on his earlier work
\cite{Simon55} for the explanation of Zipf's law in word frequency, he
assumed Gibrat's law (in a much weaker form than ours) with a boundary
condition for entry and exit of firms. In conformity with preceding
work by Champernowne for personal income \cite{Champer53}, Simon could
show that the emergence of power-law behavior is quite robust
irrespectively of modification of the stochastic process (see
\cite{SKEW} for collection of related papers). Simon modeled the
process of entry corresponding to new firms which compete with
existing firms to catch market opportunities. This line of models was
followed by \cite{Sutton95} which relaxes the assumption of Gibrat's
law, and also by \cite{Bottazzi03} which explained the Laplace
distribution for growth rate. Simon also extended his model
incorporating merger and acquisition process (see also
\cite{McLoughan95} for recent work).  These works attempted to take
into account the direct and indirect interactions among firms, which
was ignored in the assumption (c) above.

Our work is in affinity with Simon's view in the points that the
upper-tail of size distribution, Pareto-Zipf law, is focused rather
than the log-normal regime, and that the origin of it is related to
Gibrat's law and boundary condition of entry-exit of firms. It is
interesting to point out that Mansfield \cite{Mansfield62}, following
Simon's model, empirically showed that the Gibrat's law seemed to hold
only above a certain minimum size of firms. (See also \cite{Sutton97}
for the influence of \cite{Mansfield62} onto later work.)

At the end of this brief survey, let us point out why recent advent of
econophysics can have important impact on economics. The econophysics
approach attempts to treat the whole industrial organization as a
complex system, in which firms are interacting atoms, that exhibits
universal scaling laws \cite{Stanley00}.

Concerning firm size, the Pareto-Zipf power-law distribution has a
long history since the seminal work by Herbert Simon, but its study
extending to the details of growth rate was only recently facilitated
by modern datasets with good abundance and quality. In this line of
research, resent findings (e.g. \cite{Axtell01,Gaffeo03}) showed that
power-law distribution gives a very good fit for different samples of
firm size. In this paper we shall not only confirm this fact with
different European countries and for different measures of size, but
also uncover the underlying kinematics that relates Pareto-Zipf law to
Gibrat law explicitly. Following the notion of self-organized
criticality \cite{Bak96,Bak02}, the occurrence of a power-law reveals
that a deep interaction among system's subunits, reacting to
idiosyncratic shocks, leads to a critical state in which no attractive
points nor states emerge. Such interaction and critical states are so
important notions with that of self-organized criticality. Under
economic point of view, interaction means that it is not possible to
define a representative agent because the dynamics of the system is
originated just from the interaction among heterogeneous agents.
Moreover, in consequence of critical state, equilibrium exists only as
asymptote, along which the system moves from an unstable critical
point to another. The authors believe that economics can enjoy these
ideas coming from econophysics on heterogeneous interacting agents
(see \cite{Gabaix02}\cite{Gallegati03} for an example).

\section{Dataset of European Firms}

We use the dataset, Bureau van Dijk's AMADEUS, which contains
descriptive and balance data of about 260,000 firms of 45 European
countries for the years 1992--2001. For every firm are reported a
number of juridical, historical and descriptive data (as e.g.
year of inclusion, participations, mergers and acquisitions, names of
the board directors, news, etc.) and a series of data drawn from its
balance and normalized. It reports the current values (for several
currencies) of stocktaking, balance sheet (BS), profit and loss
account (P/L) and ratios. The descriptive data are frequently updated
while the numerical ones are taken from the last available balance.
Since balance year does not always match conventional year, the number
of firms included may vary during the year if one of the excluded
firms in last recording satisfy one of the criteria described below.
The amount and the completeness of available data differs from country
to country. To be included in the data set firms must satisfy at least
one of these three dimensional criteria:
\begin{itemize}
\item for U.K., France, Germany, Italy, Russian Federation and Ukraine
  \begin{itemize}
  \item operating revenue equal to at least 15 million Euro
  \item total assets equal to at least 30 million Euro
  \item number of employees equal to at least 150
  \end{itemize}
\item for the other countries
  \begin{itemize}
  \item operating revenue equal to at least 10 million Euro
  \item total assets equal to at least 20 million Euro
  \item number of employees equal to at least 100
  \end{itemize}
\end{itemize}

As a proxy for firm size, we utilize one of the financial and
fundamental variables; total-assets, sales and number of employees. We
use number of employees as a complementary variable so as to check the
validity and robustness of our results. Note that the dataset includes
firms with smaller total-assets, simply because either the number of
employees or the operating revenue (or both of them) exceeds the
corresponding threshold. We thus focus on complete sets of those firms
that have larger amount of total-assets than the threshold, and
similarly those for number of employees. For sales, we assume that our
dataset is nearly complete since a firm with a small amount of
total-assets and a small number of employees is unlikely to make a
large amount of sales. For our purposes, therefore, we discard all the
data below each corresponding threshold for each measure of firm size.
This procedure makes the number of data points much less. However, for
a several developed countries, we have enough amount of data for the study
of Gibrat's law. In what follows, our results are shown for UK and
France, although we obtained similar results for other developed
countries. The threshold for total-assets in these two countries is 30
million euros, and that for number of employees is 150 persons, as
described above. For sales, we used 15 million euros per year as a
threshold. We will also show results for Italy and Spain in addition
to U.K. and France only when examining the annual change of Pareto
indices.

It should be remarked that other problems in treating these data takes
origin from the omission, in the on-line dataset of AMADEUS, of the
date of upgrade, so that it is often not clear when a firm changed its
juridical status, or went bankrupted or inactive. For some countries
the indication activity/inactivity is not shown at all, so that it was
impossible, even indirectly, to individuate the year of exit.
Therefore, our study should be taken as the analysis conditional on
survival of firms.


\section{Firm Growth}

In this section, our results are shown for UK and France, and for
total-assets, number of employees and sales. Each list of firms is
exhaustive in the way we described in the preceding section.

\subsection{Pareto-Zipf distribution}
\label{sec:pareto}

\begin{figure}
  \centering
  \includegraphics[width=.6\columnwidth]{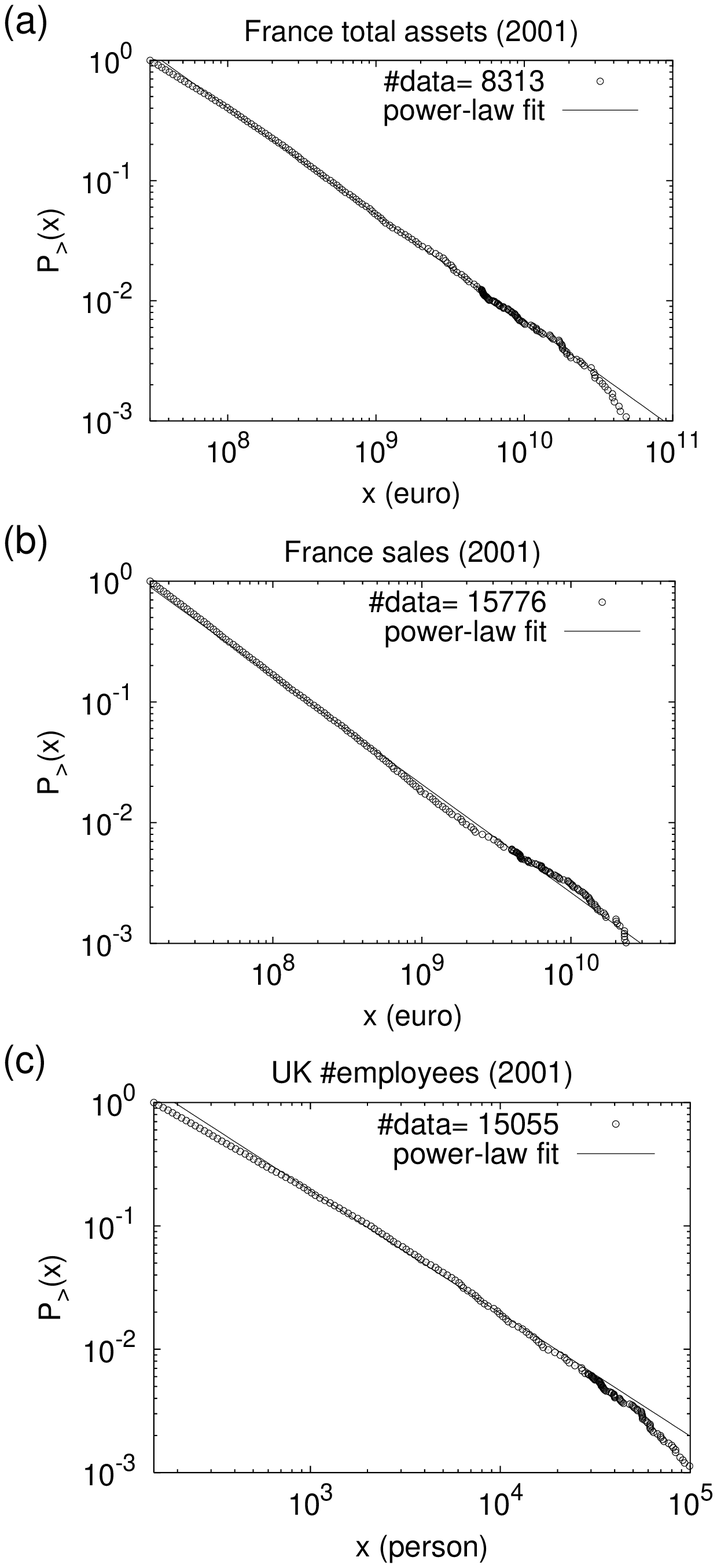}
  \caption[]{%
    Cumulative probability distribution $P_>(x)$ for firm size $x$.
    (a) Total-assets in France (2001) greater than 30 million euros,
    (b) sales in France (2001) greater than 15 million euros, (c)
    number of employees in UK (2001) larger than 150 persons. Lines
    are power-law fits with Pareto indices, (a) 0.886, (b)
    0.896, (c) 0.995 (least-square-fit in logarithmic scale).}
  \label{fig:zipf}
\end{figure}

First we show that the distribution of firm size obeys a power-law in
the range of our observation whatever we take as a variable for firm
size. \figref{fig:zipf} depicts the cumulative distributions for
total-assets in France (a), sales in France (b), and number of
employees in UK (c). The number of data points is respectively (a)
8313, (b) 15776 and (c) 15055.

Pareto-Zipf law states that the cumulative distribution $P_>(x)$ for
firm size $x$ follows \eqref{pareto}. The power-law fit for $x\geq
x_0$, where $x_0$ denotes the threshold mentioned above for each
measure of firm size, gives the values of $\mu$; (a) 0.886$\pm$0.005,
(b) 0.896$\pm$0.011, (c) 0.995$\pm$0.013 (standard error at 99\%
significance level). $\mu$ is close to unity. Note that the power-law
fit is quite well nearly three orders of magnitude in size of firms.

\begin{figure}
  \centering
  \includegraphics[width=.95\columnwidth]{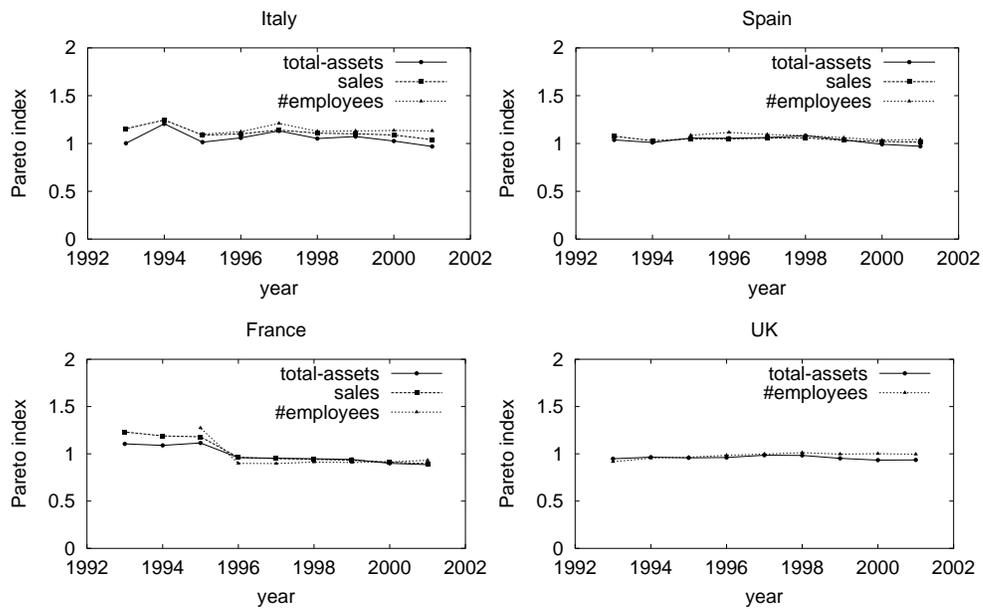}
  \caption[]{%
    Annual change of Pareto indices for Italy, Spain, France and U.K.
    from 1993 to 2001 for total-assets, number of employees, and sales
    (except U.K.). The estimate of Pareto index in each year was done
    by extracting a range of distribution corresponding to large-size
    firms, which is common to different countries but different for
    different measure of size, and by least-square-fit in logarithmic
    scales of rank and size.}
  \label{fig:euro.mu}
\end{figure}

Pareto index is surprising stable in its value. \figref{fig:euro.mu}
is a panel for the annual change of Pareto indices for four countries,
Italy, Spain, France and U.K. estimated from total-assets, number of
employees and sales (except U.K.). Different measures of firm size
give reasonably same behavior. It is observed that the value
$\mu$ is quite stable being close to unity in all the countries.

\subsection{Gibrat's law}

Let us denote a firm's size by $x$ and its two values at two
successive points in time ({\it i.e.}, two consecutive years) by $x_1$
and $x_2$. Growth rate is given by $R\equiv x_2/x_1$. We also express
the rate in terms of its logarithm, $r\equiv\log_{10}R$. We examine
the probability density for the growth rate $P(r|x_1)$ on the
condition that the firm size $x_1$ in an initial year is fixed.

For the conditioning we divide the range of $x_1$ into logarithmically
equal bins. For the total-assets in the dataset
(\figref{fig:gibrat}~(a)), the bins are taken as $x_1\in
3\times[10^{7+0.4(n-1)},10^{7+0.4n}]$ (euros) with $n=1,\cdots,5$. For
the sales in (b), $x_1\in 1.5\times[10^{7+0.4(n-1)},10^{7+0.4n}]$
(euros) with $n=1,\cdots,5$. For the number of employees in (c),
$x_1\in 1.5\times[10^{2+0.4(n-1)},10^{2+0.4n}]$ (persons) with
$n=1,\cdots,5$. In all the cases, the range of conditioning covers two
orders of magnitude in each variable.  We calculated the probability
density function for $r$ for each bin, and checked the statistical
dependence on $x_1$ by graphical method.

\figref{fig:gibrat} is the probability density function $P(r|x_1)$ for
each case. It should be noted that due to the limit $x_1 >
x_0$ and $x_2 > x_0$, the data for large negative
growth are not available. In all the cases, it is obvious that the
function $P(r|x_1)$ has little statistical dependence on $x_1$, since
all the curves for different $n$ collapse on a single curve. This
means that the growth rate is independent of firm size in the initial
year. That is, Gibrat's law holds.

\subsection{Time-reversal symmetry}

The validity of Gibrat's law in the Pareto-Zipf regime appears to be
in disagreement with recent literature on firm growth. 
In the next section, we will show that this is not actually the case
by proving that Gibrat and Pareto-Zipf are equivalent under an
assumption. The assumption is detailed balance, whose validity is
checked here.

\begin{figure}
  \centering
  \includegraphics[width=.5\columnwidth]{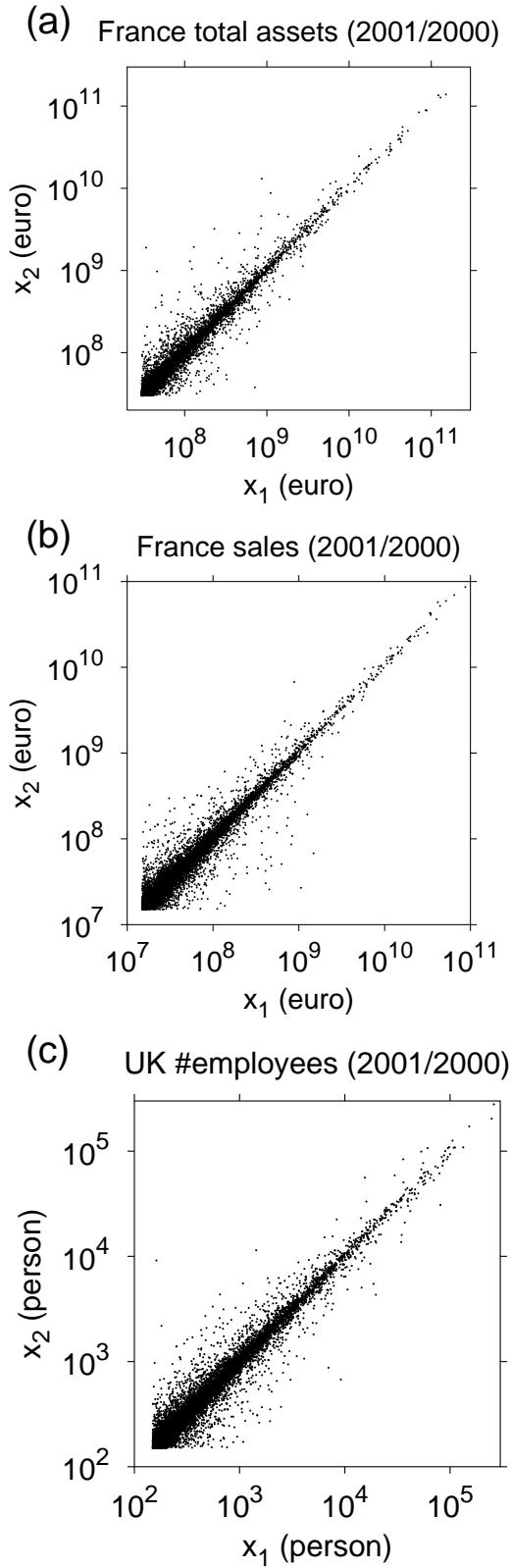}
  \caption[]{%
    Scatter-plot of all firms whose size exceeds a threshold.
    The datasets in (a)--(c) are the same as in \eqref{fig:zipf}.
    Thresholds are (a) 30 million euros for total-assets, (b) 15
    million euros for sales, and (c) 150 persons for number of
    employees. The number of such large firms is respectively 
    (a) 6969, (b) 13099 and (c) 12716.}
  \label{fig:trs}
\end{figure}

Let us denote the joint probability distribution function
for the variable $x_1$ and $x_2$ by $P_{12}(x_1,x_2)$.
The detailed balance, or what we call time-reversal symmetry, is the
assumption that $P_{12}(x_1,x_2)=P_{12}(x_2,x_1)$. The joint
probabilities for our datasets are depicted in \figref{fig:trs} as
scatter-plots of individual firms.

We used two different methods to check the validity of time-reversal
symmetry. One is an indirect way to check a non-trivial relationship
between the growth-rate in positive side ($r>0$) and that in negative
($r<0$). That is, as we shall prove in the next section, the
probability density distribution in positive and negative growth rates
must satisfy the relation given by \eqref{nocontra}, if the property
of time-reversal symmetry holds. We fitted the cumulative distribution
only for positive growth rate by a non-linear function, converted to
density function, and predicted the form of distribution for negative
growth rate by \eqref{nocontra} so as to compare with the actual
observation (see Appendix for details). In each plot of
\figref{fig:gibrat}, a solid line in the $r>0$ side is such a fit, and
a broken line in the $r<0$ side is our prediction. The agreement with
the actual observation is quite satisfactory, thereby supporting the
validity of time-reversal symmetry.

The other way we took is a direct statistical test for the symmetry in
the two arguments of $P_{12}(x_1,x_2)$. This can be done by
two-dimensional Kolmogorov-Smirnov (K-S) test, which is not widely
known but was developed by astrophysicists
\cite{NRC,Peacock83,Fasano87}. This statistical test is not strictly
non-parametric (like the well-known one-dimensional K-S test), but has
little dependence on parent distribution except through coefficient of
correlation. We compare the scatter-plot sample for $P_{12}(x_1,x_2)$
with another sample for $x_1$ and $x_2$ interchanged by making the
null hypothesis that these two samples are taken from a same parent
distribution.  We used the logarithms $\xi_1=\log x_1$ and $\xi_2=\log
x_2$, and added constants to $\xi_1$ and $\xi_2$ so that the average
growth rate is zero. This addition (or multiplication in
$x_1$ and $x_2$) is simply subtracting the nominal effects due to
inflation, etc.  We applied two-dimensional K-S test to the resulting
samples.  The null hypothesis is not rejected in 95\% significance
level in all the cases we studied.

\section{Pareto-Zipf's law and Gibrat's law under detailed balance}

In the preceding section, we have shown that both of Pareto-Zipf and
Gibrat's laws hold for large firms. This suggests that these two laws
are closely related with each other. We show in this section that in
fact they are equivalent to each other under the condition of detailed
balance.

Let $x$ be a firm's size, and let its two values at two successive
points in time ({\it i.e.}, two consecutive years) be denoted by $x_1$
and $x_2$.  We denote the joint probability distribution function
(pdf) for the variable $x_1$ and $x_2$ by $P_{12}(x_1,x_2)$. The
joint pdf of $x_1$ and the growth rate $R=x_2/x_1$ is denoted by
$P_{1R}(x_1,R)$.  Since $P_{12}(x_1,x_2)dx_1 dx_2=P_{1R}(x_1,R)dx_1
dR$ under the change of variables from $(x_1,x_2)$ to $(x_1,R)$, these
two pdf's are related to each other as follows:
\begin{equation}
  P_{1R}\left(x_1,{x_2\over x_1}\right)=x_1 P_{12}(x_1,x_2) .
  \label{trans}
\end{equation}

We define conditional probabilities:
\begin{eqnarray}
  P_{1R}(x_1,R)&=&P_1(x_1)\,Q(R\,|\,x_1)  \label{defQ}\\
  &=&P_R(R)\,S(x_1|\,R),   \label{defS}
\end{eqnarray}
Both $P_1(x_1)$ and $P_R(R)$ are marginal:
%
\begin{eqnarray}
  P_1(x_1)&=&\int_0^\infty P_{1R}(x_1,R)dR
  \left( = \int_0^\infty P_{12}(x_1, x_2)dx_2\right),  \label{defP1}\\
  P_R(R)&=&\int_0^\infty P_{1R}(x_1,R)dx_1,   \label{defPR}
\end{eqnarray}
since the following normalizability conditions are satisfied:
\begin{eqnarray}
  1&=&\int_0^\infty Q(R|x_1)dR ,\label{normQ}\\
  1&=&\int_0^\infty S(x_1|R)dx_1 .\label{normS}
\end{eqnarray}

Three phenomenological properties can be summarized as follows.
\begin{enumerate}
\item[(A)] {\sl Detailed Balance (Time-reversal symmetry)}:\\
The joint pdf $P_{12}(x_1, x_2)$ is a symmetric function:
\begin{equation}
P_{12}(x_1, x_2)=P_{12}(x_2, x_1).
\label{Tinv}
\end{equation}
\item[(B)] {\sl Pareto-Zipf's law:}\\
The pdf $P_1(x)$ obeys power-law for large $x$:
\begin{equation}
P_1(x)\propto x^{-\mu-1},
\label{originalPareto}
\end{equation}
for $x \rightarrow \infty$ with $\mu > 0$.
\item[(C)] {\sl Gibrat's law}:\\
The conditional probability $Q(R\,|\,x)$ is independent of $x$:
\begin{equation}
Q(R\,|\,x)= Q(R).
\label{univ}
\end{equation}
We note here that this holds only for large $x$, because we confirmed
it in actual data only in that region, and because otherwise it leads
to an inconsistency, as we will see shortly.  This relation was called
{\sl Universality} in \cite{Yoshi03a,Aoyama03a,Yoshi03b,Aoyama03b}.
All the arguments below is restricted in this region.
\end{enumerate}

Before starting our discussion of interrelation between these
properties, let us first rewrite the detailed balance condition (A) in
terms of $P_{1R}(x_1,R)$:
\begin{eqnarray}
  P_{1R}(x_1,R)&=&x_1 P_{12}(x_1,x_2) \nonumber\\
  &=&x_1 P_{12}(x_2,x_1) \nonumber\\
  &=&{x_1\over x_2} x_2 P_{12}(x_2,x_1) \nonumber\\
  &=&R^{-1}P_{1R}\left(x_2, R^{-1}\right) , \label{eqa}
\end{eqnarray}
where \eqref{Tinv} was used in the second line, and \eqref{trans} was
used in the first and the third line. The above relation may be
rewritten as follows by the use of the conditional probability
$Q(R\,|\,x_1)$ in \eqref{defS};
\begin{eqnarray}
  {Q(R^{-1}\,|\,x_2)\over Q(R\,|\,x_1)}=R
  {P_1(x_1)\over P_1(x_2)} .
\label{usethis}
\end{eqnarray}

In passing, it should be noted that \eqref{eqa} leads to the
following:
\begin{eqnarray}
  P_R(R)&=&\int_0^\infty P_{1R}(x_1,R)dx_1 \nonumber\\
  &=&\int_0^\infty R^{-1} P_{1R}\left(x_2, R^{-1}\right)dx_1 \nonumber\\
  &=&\int_0^\infty R^{-2} P_{1R}\left(x_2, R^{-1}\right)dx_2 \nonumber\\
  &=& R^{-2}P_R\left(R^{-1}\right) \label{invPR}
\end{eqnarray}
where \eqref{eqa} was used in the second line, and the third line is
merely change of integration variable.  This relation between the
marginal growth-rate pdf $P_R(R)$ for positive growth ($R>1$) and
negative growth ($R<1$) leads to the following relation, as it should:
\begin{equation}
\int_1^\infty P_R(R)dR = \int_0^1 P_R(R)dR.
\end{equation}

\subsection{(A)+(C)$\rightarrow$(B)}

Let us first prove that the properties (A) and (C) lead to (B). By
substituting the Gibrat's law \eqref{univ} in \eqref{usethis}, we
find the following:
\begin{equation}
 {P_1(x_1)\over P_1(x_2)}= \frac1R \,{Q(R^{-1})\over Q(R)}.    
\label{usethis2}
\end{equation}
This relation can be satisfied only by a power-law function
\eqref{originalPareto}.

\noindent [Proof]\\
Let us rewrite \eqref{usethis2} as follows:
\begin{equation}
  P_1(x)=G(R)P_1(Rx),
\end{equation}
where $x$ denotes $x_1$, and $G(R)$ denotes the right-hand side of
\eqref{usethis2}, i.e.
\begin{equation}
  G(R)\equiv{1\over R}{Q(R^{-1})\over Q(R)}.
  \label{grdef}
\end{equation}
We expand this equation around $R=1$ by denoting
$R=1+\epsilon$ with $\epsilon \ll 1$ as
\begin{eqnarray}
  P_1(x)&=&G(1+\epsilon) P_1((1+\epsilon)x)
  \nonumber\\
  &=& (1 + G'(1)\epsilon+\cdots)(P_1(x)+P_1'(x)\epsilon x +\cdots)
  \nonumber\\
  &=& P_1(x) + \epsilon (G'(1) P_1(x)+ x P_1'(x)) + O(\epsilon^2),
\end{eqnarray}
where we used the fact that $G(1)=1$. We also assumed that the
derivatives $G'(1)$ and $P_1'(x)$ exists in the above, whose validity
should be checked against the results. From the above, we find that
the following should be satisfied
\begin{equation}
  G'(1) P_1(x)+ x P_1'(x)=0,
  \label{diffx}
\end{equation}
whose solution is given by
\begin{equation}
  P_1(x)=C x^{-G'(1)}.
  \label{proofres}
\end{equation}
This is the desired result, Pareto-Zipf's law, and is consistent with
the assumption made earlier that $P_1'(x)$ exists. By substituting the
result \eqref{proofres} in \eqref{grdef} and \eqref{usethis2}, we find
that
\begin{equation}
  G(R)=R^{G'(1)},
  \label{grrgone}
\end{equation}
which is consistent with the assumption that $G'(1)$ exists.\\
\rightline{[Q.E.D.]}

From \eqref{grdef} we may calculate $G'(1)$ in terms of derivatives of
$Q(R)$. It should, however, be noted that $Q(R)$ has a cusp at $R=1$
as is apparent in \figref{fig:gibrat}, and therefore $Q'(R)$ is
expected not to be continuous at $R=1$. Bearing this in mind, we
calculate $G(1+\epsilon)$ for $0 < \epsilon \ll 1$ as follows:
\begin{eqnarray}
  G(1+\epsilon)&\simeq&
  \frac1{1+\epsilon}\, \frac{\,Q(1-\epsilon)\,}{Q(1+\epsilon)}
  \nonumber\\
  &\simeq& (1-\epsilon)\,
  \frac{\,Q(1)-\epsilon \,Q^{-\prime}(1)\,}{Q(1)+\epsilon \,Q^{+\prime}(1)}
  \nonumber\\
  &\simeq& 
  G(1) +\epsilon \left(-1-\frac{Q^{+\prime}(1)+Q^{-\prime}(1)}{Q(1)}\right),
\end{eqnarray}
where we denoted the right-derivative and left-derivative of
$Q(R)$ at $R=1$ by the signs $+$ and $-$ in the superscript,
respectively.  From the above, we find that
\begin{equation}
  G'(1)=-1-\frac{Q^{+\prime}(1)+Q^{-\prime}(1)}{Q(1)},
  \label{tasuQQ}
\end{equation}
From \eqref{proofres} and \eqref{tasuQQ}, we find that
\begin{equation}
  \frac{Q^{+\prime}(1)+Q^{-\prime}(1)}{Q(1)}=-\mu-2.
  \label{Qdiffe}
\end{equation}

From eqs.(\ref{grdef}) and (\ref{grrgone}), we find the following
relation:
\begin{equation}
  Q(R)=R^{-\mu-2} Q(R^{-1}),
  \label{nocontra}
\end{equation}
which should be in contrast to \eqref{invPR}. This is related to the
point that we mentioned in \eqref{univ}: If the Gibrat's law
\eqref{univ} holds for all $x\in [0,\infty]$, then $P_R(R)=Q(R)$ from
\eqref{defPR}. If so, \eqref{nocontra} contradicts to \eqref{invPR}
since $\mu >0$. Besides, the Pareto-Zipf's law we derived from
Gibrat's law is not normalizable if it holds for any $x$.  Therefore,
Gibrat's law should hold only for large $x$.

The result \eqref{nocontra} shows that the function $Q(R)$ is
continuous at $R=1$, as is easily seen by substituting $R=1+\epsilon$
with $\epsilon >0$ on both hand side and taking the limit $\epsilon
\rightarrow +0$.  Also, by taking the derivative of the both hand side
and taking the limit in a similar manner, we can reproduce
\eqref{Qdiffe}.


\subsection{(A)+(B)$\rightarrow$ ?}

Let us next examine what we obtain if we had only Pareto-Zipf's law
instead of Gibrat's law under the detailed balance.

In this case, substituting the Pareto-Zipf's law
\eqref{originalPareto}  into \eqref{usethis} we find that
\begin{equation}
  {Q(R^{-1}\,|\,Rx)\over Q(R\,|\,x)}=R^{\mu+2},
  \label{gr1}
\end{equation}
where we denote $x_1$ by $x$ and $x_2$ by $Rx$.
We now define a function $H(z,x)$ as
\begin{equation}
  Q(R\,|\,x)=x^{\mu+2}H(R^{1/2}x, x).
  \label{gr2}
\end{equation}
It should be noted that this does not constrain $Q(R\,|\,x)$ in any way:
arbitrary function of the variable 
$R$ and $x$ can be written in the form of \eqref{gr2}.
By substituting \eqref{gr2} into \eqref{gr1}, we find that
\begin{equation}
  H(R^{1/2}x, Rx)=H(R^{1/2}x, x),
\end{equation}
which means that the function $H(z,x)$
has the following invariance property.
\begin{equation}
  H(z,x)=H(z,z^2/x).
  \label{gr3}
\end{equation}
Other than this constraint and some trivial constraint such as continuity, 
there is no nontrivial constraint on $H(z,x)$ or $Q(R\,|\,x)$.

The results eqs.~(\ref{gr2}) and (\ref{gr3}) is a generalization of
the property \eqref{nocontra} we found earlier \cite{Yoshi03a}. In
fact, the property \eqref{nocontra} follows from \eqref{gr3} in the
special case:
\begin{equation}
  H(z,x)=Q((z/x)^2)\,x^{-\mu-2} ,
  \label{grspec}
\end{equation}
for which \eqref{gr2} becomes \eqref{univ}, namely the statement of
Gibrat's law.


\subsection{(B)+(C)$\rightarrow$(A)?}

Let us discuss the last question: Under Pareto's and Gibrat's laws,
what can we say about the detailed balance? In order to answer this, we
use \eqref{originalPareto} and \eqref{univ} to write $P_{1R}(x,R)$ for
large $x$ as follows:
\begin{equation}
  P_{1R}(x,R)=A x^{-\mu-1} Q(R),
  \label{PG}
\end{equation}
where $A$ is a proportionality constant. According to \eqref{eqa},
the detailed balance is satisfied if this is equal to
\begin{equation}
  R^{-1}P_{1R}(xR,R^{-1})=A x^{-\mu-1} R^{-\mu-2}Q(R^{-1}),
\end{equation}
where we used \eqref{PG}. Therefore, we find that the detailed balance
condition is equivalent to \eqref{nocontra} in this case.

\begin{figure}
  \centering
  \includegraphics[width=.7\columnwidth]{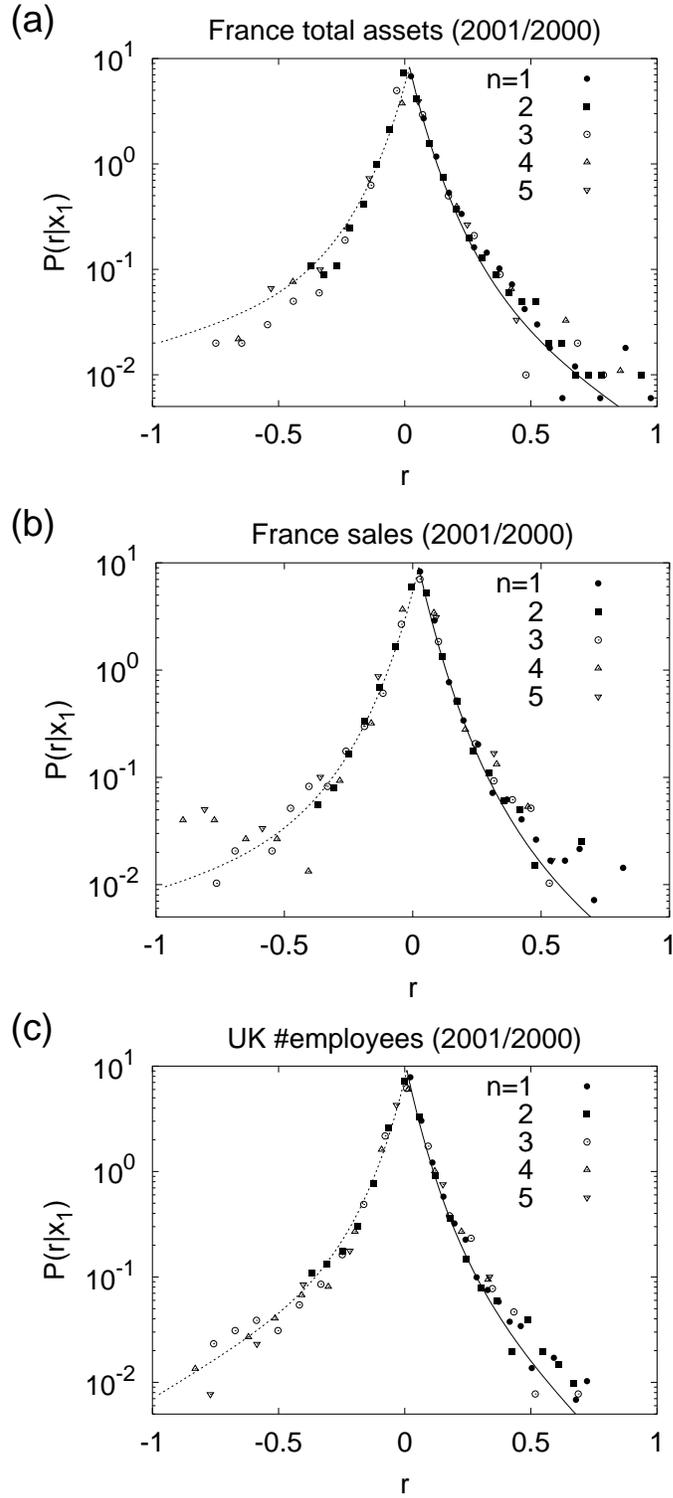}
  \caption[]{%
    Probability density $P(r|x_1)$ of growth rate
    $r\equiv\log_{10}(x_2/x_1)$ for the two years, 2000/2001. The
    datasets in (a)--(c) are the same as in \eqref{fig:zipf}.
    Different bins of initial firm size with equal magnitude in
    logarithmic scale were taken over two orders of magnitude as
    described in the main text.
    The solid line in the portion of positive growth ($r>0$)
    is a non-linear fit. The dashed line ($r<0$) in the negative side
    is calculated from the fit by the relation given in the equation
    \eqref{nocontra}.}
  \label{fig:gibrat}
\end{figure}

Summarizing this section, we have proved that under the condition of
detailed balance (A), if the Pareto-Zipf law (B) holds in a region of firm
size, then the Gibrat's law (C) must hold in the region, and {\it vice
 versa}. The condition (A) means detailed-balance. On the other hand,
if both of (B) and (C) hold, (A) follows provided that
\eqref{nocontra} holds. \eqref{nocontra} is our prediction which gives
a non-trivial relation between positive growth ($R>1$) and negative
($R<1$). This kinematic relation was empirically verified in
\figref{fig:gibrat}. See also previous work
\cite{Yoshi03a,Aoyama03a,Yoshi03b,Aoyama03b} for the validity of this relation in
personal income and firms tax-income in Japan.

\section{Summary}

The distribution of firm size is quite often dominated by power-law in
the upper tail over several orders of magnitude. This regime of
Pareto-Zipf law is different from log-normal distribution in the lower
and sometimes wider regime of firm size. The upper tail is occupied by
a small number of firms, but they dominate a large fraction of total
sum of firm size.

By using exhaustive datasets of those large firms and with different
measures of firm size in Europe, we show that the Pareto-Zipf law
holds as in \eqref{pareto} for firm size $x$ larger than observational
threshold $x_0$, and that Gibrat's law of proportionate effect holds
as in \eqref{gibrat} for successive sizes $x_1$ and $x_2$ exceeding
$x_0$, stating that the growth rate of each firm is independent of
initial size. We also find that detailed balance holds which means
that the frequency of transition from $x_1$ to $x_2$ is statistically
the same as that for its reverse process. The Gibrat's law,
Pareto-Zipf's law and detailed balance condition are related to each
other. We prove various relationships among them.  It follows as one
of the consequences that there exists a relation between the positive
and negative sides of the distribution of growth rate via the Pareto
index. The relation is confirmed empirically in our dataset of
European firms.

\begin{ack}
  We thank Edmondo Di Giuseppe for helping the preparation of raw
  datasets before analysis. Y.~F. and W.~S. thank Katsunori Shimohara
  for encouragement. Supported in part by grants from the
  Telecommunications Advancement Organization in Japan and from Japan
  Association for Cultural Exchange.
\end{ack}


\clearpage
\clearpage
\clearpage
\clearpage


\clearpage
\appendix
\section{Fitting distribution of growth rate}

For the purpose of fitting probability density function of
positive growth rate ($R>1$), we used {\it cumulative} distribution of
positive growth rate, defined by
\[
  P_>^+(R)=\mbox{Prob}\{\mbox{\boldmath$R$}>R|R>1\} .
\]
$P_>^+(R)$ can be estimated, as usual, by size versus rank plot
restricted only for $R>1$ as follows. Let the number of all firms with
$R>1$ be $N_+$, and sort their growth rates in descending order:
$R^{(1)}\geq R^{(2)}\geq\cdots\geq R^{(k)}\geq\cdots\geq R^{(N_+)}$.
Then the estimate is given by
\[
  P_>^+(R)={k\over N_+}=C^{-1}\int_{V^{(k)}}P_{1R}(x_1,R)dx_1 dR ,
\]
where $V^{(k)}=\{(x_1,R)|x_1\geq x_0,R\geq R^{(k)}\}$ ($x_0$ is the
observational threshold mentioned in section \ref{sec:pareto}), and
$C$ is the normalization:
\[
  C=\int_{x_0}^\infty dx_1 \int_1^\infty dR P_{1R}(x_1,R) .
\]
Using the observational fact that \eqref{univ} holds in the region
$\{(x_1,R)|x_1\geq x_0,R\geq1\}$, the above equation for $P_>^+(R)$
reads
\begin{equation}
  P_>^+(R)=C^{-1}\int_{x_0}^\infty dx_1 P_1(x_1)dx_1
    \int_R^\infty dR' Q(R')
  =Q_0{}^{-1}\int_R^\infty Q(R')dR' ,
  \label{cumplus}
\end{equation}
where the normalization factor is written by
\begin{equation}
  Q_0=\int_1^\infty Q(R')dR' .
  \label{Q0}
\end{equation}
By taking derivative of \eqref{cumplus} with respect to $R$, it
follows that
\begin{equation}
  Q(R)=-Q_0{d\over dR}P_>^+(R) .
  \label{cumplusQ}
\end{equation}

\begin{figure}
  \centering
  \includegraphics[width=.7\columnwidth]{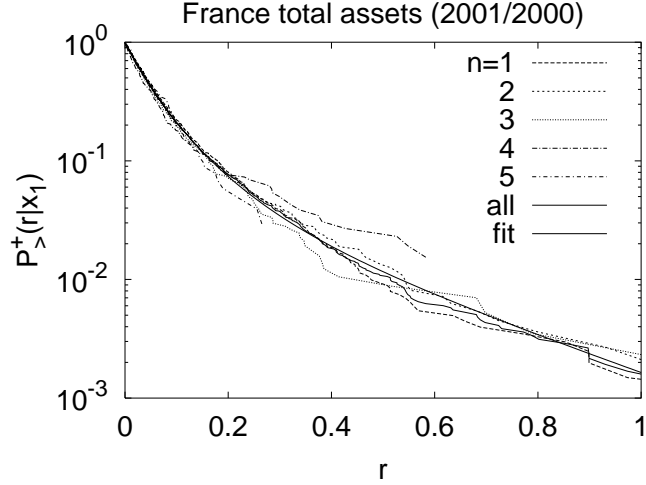}
  \caption[]{%
    Cumulative probability $P_>^+(R=10^r)$ for the growth of
    total-assets in France (2001/2000). $n$ is the index of bin used
    in \figref{fig:gibrat}~(a), ``all'' means the plot for
    all the dataset of positive $r$. ``Fit'' is done by the non-linear
    function given by the equation \eqref{cum-r-fit}.}
  \label{fig:cum-r}
\end{figure}

We empirically found that the rank-size plot can be well fitted by a
non-linear function of the form:
\begin{equation}
  \log_{10}P_>^+(R=10^r)=-a(1-e^{-br})-c r \equiv F(r) ,
  \label{cum-r-fit}
\end{equation}
where $a$, $b$ and $c$ are parameters. An example is given in
\figref{fig:cum-r} for France total-assets (2001/2000). Cumulative
probabilities $P_>^+(r|x_1)$ (the left-hand side of \eqref{cum-r-fit})
conditioned on an initial year's total-assets are shown for each of
the same bins used in \figref{fig:gibrat}~(a), but restricted to the
data with positive $r$. The non-linear fit done by \eqref{cum-r-fit} is
represented by a solid and bold line in the figure. Note also that the
curves for different bins almost collapse because of the statistical
independence of $x_1$.

Under the change of variable, $r=\log_{10}R$, the probability density
for $r$ defined by $q(r)$ is related to that for $R$ by
\begin{equation}
  \log_{10}q(r)=\log_{10}Q(R=10^r)+r+\log_{10}(\ln 10)
  \label{qandQ}
\end{equation}
Therefore it follows from \eqref{cumplusQ} and \eqref{qandQ} that
\begin{equation}
  \log_{10}q(r)=F(r)+\log_{10}\left[-{dF(r)\over dr}\right]
  +\log_{10}Q_0+\log_{10}(\ln 10) .
  \label{fit-qr}
\end{equation}
In each plot of \figref{fig:gibrat}, the solid curve is given by
\eqref{fit-qr}, where $P(r|x_1)$ denotes the probability density
function $q(r)$ for $r$, conditioned on an initial year's size $x_1$.

The relation \eqref{nocontra} for positive ($R>1$) and negative
($R<1$) growth rates can be written in terms of $q(r)$ as
\begin{equation}
  \log_{10}q(r)=-\mu r+\log_{10}q(-r) ,
  \label{inv-qr}
\end{equation}
which is easily shown by \eqref{qandQ}. 
In each plot of \figref{fig:gibrat}, the dotted curve for negative
growth rate ($r<0$) is obtained from the solid curve for positive one
($r>0$) through the relation \eqref{inv-qr}.

\end{document}